\documentclass{ws-p8-50x6-00}
\newcommand{\trho}{\tilde{\rho}}
\newcommand{\cum}[1]{\left\langle\,#1\,
\right\rangle_{\Omega,x_{\min},c}^{x_a,x_b}}
\begin{document}

\title{Smearing Formulas for Density Matrices}

\author{M. Bachmann, H. Kleinert, and A. Pelster}

\address{Institut f\"ur Theoretische Physik, Freie Universit\"at Berlin, \\Arnimallee 14, 14195 Berlin, Germany}

\maketitle

\abstracts{
We report on the development of a
systematic variational perturbation theory for the
euclidean path integral representation of the density matrix
based on new smearing formulas for harmonic correlation functions.
As a first application, we present
the lowest-order approximation for the
radial distribution function of an electron in a hydrogen atom.
}
\section{Introduction}
Quantum statistical free energies can be calculated
 to any desired accuracy with the help of variational perturbation theory~\cite{kl0,kl1}.
This is a systematic generalization of an earlier
 rough variational approach~\cite{fk,gt}
based on the Jensen-Peierls inequality.
The approximation consists of an optimized expansion,
in which each term can be expressed via a simple smearing formula~\cite{kkp}
applied to powers of the interaction, which can
be polynomial as well as nonpolynomial.

In Sect.~\ref{dmsect} we
sketch the extension of this theory
to density matrices, thus enabling us to calculate
 very accurately
local statistical
properties of a quantum mechanical system at all temperatures and coupling
strengths. The corresponding smearing formulas
are derived in Sect.~\ref{smsect} and applied in Sect.~\ref{appsect} to obtain
the lowest-order approximation for the
density distribution of a Coulomb system.
\section{Variational Perturbation Theory for Density Matrices}
\label{dmsect}
Variational perturbation expansions
for the free energy approximate
 arbitrary quantum statistical
systems by  optimized local
perturbation expansions around harmonic systems
with different trial frequencies.
The optimization is performed
separately
for each path average $x_0=\int _0^{\hbar  \beta }x(\tau )/\hbar\beta$.
This
ensures
a rapid convergence of such expansions at higher temperatures
by removing
the
fluctuations of $x_0$
which diverge linearly with the temperature.
The fluctuations of  $x_0$
 are accounted for
at the end by an ordinary integral.
Since these fluctuations
probe, at higher temperatures,
the entire potential, they must be
done numerically, and the fact that this is always possible with high accuracy
is one of the reasons for the
quality of the approximations~\cite{kl1}.

When applying variational perturbation theory to
density matrices,
the special role of the
 $x_0$-fluctuations
diappears since $x_0$
remains always
close to some average of the fixed endpoints of the paths.
Then $x_0$ may be treated perturbatively~\cite{bkp}
together with the other Fourier components of the paths.
We may therefore directly look for an optimized perturbation expansion
of the path integral
for the unnormalized
density matrix
\begin{equation}
  \label{prop}
  \trho(x_a,x_b)=\int\limits_{x_a,0}^{x_b,\hbar\beta}{\cal D}x\,
\exp\left\{-\frac{1}{\hbar} {\cal A}[x] \right\},
\end{equation}
where $\beta=1/k_BT$
and
${\cal A}$ is the euclidean
 action. In order to obtain
a variational approximation, we divide the full action into a harmonic trial
action with center $x_{\min}=x_{\min}(x_a,x_b)$ and
frequency $ \Omega=\Omega(x_a,x_b,x_{\min}) $.
Denoting the trial action by
${\cal A}_{\Omega,x_{\min}}$, the remainder
${\cal A}_{\rm int}={\cal A}-{\cal A}_{\Omega,x_{\min}}=
\int_0^{\hbar\beta}d\tau\,V_{\rm int}(x(\tau))$ is
treated
as a
perturbation. The result
can be written as an exponential of a cumulant
expansion
(cumulants being indicated by subscript $c$)
\begin{eqnarray}
\label{nd}
\trho(x_a,x_b)&=&\trho_{\Omega,x_{\min}}(x_a,x_b)\nonumber\\
&\times&\exp\left\{-\frac{1}{\hbar}\cum{{\cal A}_{\rm int}[x]}+
\frac{1}{2\hbar^2}\cum{{\cal A}_{\rm int}^2[x]}-\ldots\right\}.
\end{eqnarray}
The prefactor
 $\trho_{\Omega,x_{\min}}(x_a,x_b)$
is the unnormalized density matrix of the
 displaced euclidean harmonic propagator.
The connected correlation functions
in the exponent of (\ref{nd}) consist
of
the harmonic expectation values
\begin{eqnarray}
  \label{expval}
  &&\langle\,{\cal A}^n[x]\,\rangle_{\Omega,x_{\min}}^{x_a,x_b}=
[\trho_{\Omega,x_{\min}}(x_a,x_b)]^{-1}\nonumber\\
&&\hspace{30pt}\times\int\limits_{x_a,0}^{x_b,\hbar\beta}{\cal D}x
\prod\limits_{k=1}^n\left[\int_0^{\hbar\beta}d\tau_k\,
V_{\rm int}(x(\tau_k)) \right]\exp
\left\{-\frac{1}{\hbar}{\cal A}_{\Omega,x_{\min}}[x] \right\}.
\end{eqnarray}
Truncating the series (\ref{nd}) after the $N$th term, we find
the approximation $\trho_N(x_a,x_b;\Omega,x_{\min})$ to
the euclidean propagator (\ref{prop}).
As the exact propagator $\trho(x_a,x_b)$ does not depend on the variational
parameters, we expect the best approximation
$\trho_N(x_a,x_b;\Omega,x_{\min})$
to depend
minimally
on them.
To determine
the optimal values $\Omega^N$ and
$x_{\min}^N$, we thus
solve
the extremality conditions
$\partial \trho_N(x_a,x_b;\Omega,x_{\min})/\partial \Omega=0$ and
$\partial \trho_N(x_a,x_b;\Omega,x_{\min})/\partial x_{\min}=0$.
If no extremal point is found, higher derivatives can be used~\cite{kl1}.
The associated
density matrix $\rho_N$ is found
by normalizing $\trho_N$:
\begin{equation}
  \label{dens}
  \rho_N(x_a,x_b)=\frac{\trho_N(x_a,x_b;
\Omega^N( x_a , x_b,x_{\min}(x_a,x_b) ),x_{\min}^N( x_a , x_b ))}{\int_{-\infty}^{+\infty}
dx\,\trho_N(x,x;\Omega^N( x , x,x_{\min}^N(x,x)  ),x_{\min}^N( x , x ) )}.
\end{equation}
\section{Smearing Formula}
\label{smsect}
To determine the expectation values (\ref{expval}),
we proceed as in Ref.~\cite{kkp2}, and
introduce the Fourier identity
\begin{equation}
  \label{sm1}
  {\cal A}_{\rm int}[x]=\int\limits_0^{\hbar\beta}
d\tau\,\int\limits_{-\infty}^{+\infty}dz\,V_{\rm int}(z)
\int\limits_{-\infty}^{+\infty}\frac{dk}{2\pi}\,e^{ikz}\exp
\left\{\frac{1}{\hbar}\int_0^{\hbar\beta}d\tau'\,j(k,\tau')x(\tau') \right\}
\end{equation}
with the current $j(k,\tau')=-i\hbar k\delta(\tau-\tau')$. Thus the
expectation values (\ref{expval}) are reduced to path integrals for the
euclidean harmonic propagator of exponentials of the
fluctuating variables.
But harmonic expectation values of products of exponentials
can be written as products of exponentials of pair correlation functions
as a simple generalization of Wick's rule (the simplest
well-known application of this being the calculation of the
Debye-Waller factor for harmonic phonons).
Performing the remaining
Gaussian integrals we obtain directly the smearing formula
\begin{eqnarray}
  \label{sm2}
\hspace{-1.6cm}&&\langle\,{\cal A}_{\rm int}^n[x]\,
\rangle_{\Omega,x_{\min}}^{x_a,x_b}=\prod\limits_{k=1}^n
\left[\int_0^{\hbar\beta}d\tau_k\int_{-\infty}^{+\infty}dz_k\,
V_{\rm int}(z_k+x_{\min}) \right]\nonumber\\
&&\times\frac{1}{\sqrt{(2\pi)^n{\rm det}\,G}}
 \prod_{k,l=1}^n\exp\left\{-\frac{1}{2}\,
[z_k-x_{\rm cl}(\tau_k)]\,G_{kl}^{-1}(\tau_k,\tau_l)\,[z_l-x_{\rm cl}(\tau_l)] \right\},
\end{eqnarray}
where $x_{\rm cl}(\tau)$ is the classical harmonic path and $G$
denotes the $n \times n$-matrix of harmonic Green functions
\begin{equation}
  \label{sm3}
  G_{kl}(\tau_k,\tau_l)=\frac{\hbar}{2M\Omega}\frac{\cosh\Omega(|\tau_k
-\tau_l|-\hbar\beta)-\cosh\Omega(\tau_k+\tau_l-\hbar\beta)}{\sinh\hbar
\beta\Omega}\,.
\end{equation}
An important advantage of the smearing formula (\ref{sm2}) over the
conventional diagrammatic perturbation expansions
is that it also allows to calculate correlation functions for
nonpolynomial problems.
\section{Electron Distribution in Hydrogen Atom}
\label{appsect}
\begin{figure}[t]
\epsfxsize=8cm
\centerline{\epsfbox{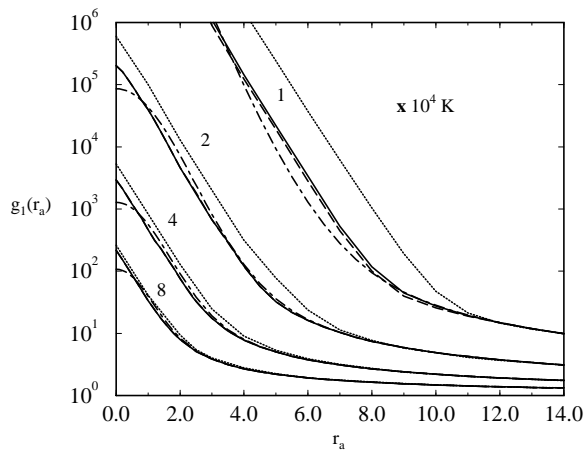}}
\caption[]{\label{fig1} Radial distribution function for an electron in a hydrogen atom. The first-order results obtained with isotropic
(dashed curves) and anisotropic (solid) variational perturbation theory are plotted against
Storer's results~\cite{storer} (dotted) and an earlier curve obtained from the first-order effective classical potential~\cite{kl1,jk}
(dash-dotted).}
\end{figure}

As an example, consider the
electron distribution in the hydrogen
atom for different temperatures shown in Fig.~\ref{fig1}.
Denoting by $r_a=|{\bf r}_a|$ the distance of the electron from the proton,
one usually plots the so-called {\em pair distribution function\/}
$g_1({\bf r}_a)$, which is  equal to $(2\pi\beta)^{3/2}$ times
 the unnormalized density $\trho_1({\bf r}_a,{\bf r}_a)$. Our result is better than an
earlier approximation obtained from a simple smearing of the effective classical
potential of the system~\cite{kl1,jk}.
\section*{Acknowledgments}
The work of one of us (M.B.) is supported by the Studienstiftung des
deutschen Volkes.

\end{document}